\documentclass[11pt]{article}

\usepackage[a4paper, total={6in, 8in}]{geometry}

\usepackage{amssymb}
\usepackage{amsmath}
\usepackage{url}

\usepackage{graphicx}
\usepackage{subfigure}
\usepackage{epstopdf}
\usepackage{wrapfig}
\usepackage{tikz,pgfplots}
\usepackage{pgfplotstable}
\usepackage{amscd,amstext,bbm, bm}

\usepackage{verbatim}

\usepackage{adjustbox}
\usepackage{booktabs} 

\usepackage{xcolor}

\usepackage{algorithm,algpseudocode}

\newcommand{\ML}{\textsc{ML}}
\newcommand{\kasandr}{\textsc{Kasandr}}
\newcommand{\BPR}{\texttt{BPR}}
\newcommand{\caser}{\texttt{Caser}}

\newcommand{\GRU}{\texttt{GRU4Rec}}

\newcommand{\mapfive}{\texttt{MAP@5}}
\newcommand{\mapten}{\texttt{MAP@10}}
\newcommand{\ndcgfive}{\texttt{NDCG@5}}
\newcommand{\ndcgten}{\texttt{NDCG@10}}
\newcommand{\SO}{\texttt{SNAPE}}
\newcommand{\MOSAIC}{\texttt{MOSAIC}}

\begin{document}
%
\title{\bf Recommender systems: when memory matters\footnote{The paper has been accepted to the 44-th European Conference on Information Retrieval (ECIR), 2022}}

%
%

\author{Aleksandra Burashnikova$^{1,3}$, Marianne Clausel$^{2}$, Massih-Reza Amini$^{3}$,\\ Yury~Maximov$^{4}$, and Nicolas Dante$^{2}$\\[2ex]
$^{1}$ Skolkovo Institute of Science and Technology, Russia\\ 
$^{2}$ University of Lorraine, France\\
$^{3}$ University Grenoble-Alpes, France\\
$^{4}$ Los Alamos National Laboratory, USA
}
%
%
%

\maketitle

\begin{abstract}
In this paper, we study the effect of long memory  in the learnability of a sequential recommender system including users implicit feedback.  We propose an online algorithm, where model parameters are updated user per user over blocks of items constituted by a sequence of unclicked items followed by a clicked one. We illustrate through thorough empirical evaluations that filtering users with respect to the degree of long memory contained in their interactions with the system allows to substantially gain in performance with respect to MAP and NDCG especially in the context of training large-scale Recommender Systems.
\end{abstract}

\section{Introduction}
Recently there has been a surge of interest in the design of personalized recommender systems (RS) that adapt to user's taste based on their {\it implicit feedback}, mostly in the form of clicks. The first works on RS assume  that users provide an {\it explicit feedback} (as scores) each time an item is shown to them. However providing a score is usually time consuming; and in many practical situations users may provide no feedback when they are not interested in some items that are shown to them, or, they may click on items that are likely to be of their interest.

In the last few years, most works were interested in taking into account the sequential nature of user/item interactions in the learning process \cite{Fang:20}. These approaches are mainly focused in the design of sequential neural networks as RNN or LSTM for predicting, in the form of posterior probabilities, the user's preference given the items \cite{zhang2019deep}.  Models from the feedback history of a given user his next positive feedback \cite{donkers2017sequential}. All these strategies consider only the sequence of positive feedback i.e.  viewed items that are clicked or purchased; and rely on the underlying assumption that user/item interactions are {\it homogeneous} and {\it persistent} in time, motivating the design of RS based on stationary neural networks with nice memory properties.

In this paper, we  put in evidence $(a)$  the effectiveness of taking into account negative feedback along with positive ones in the learning of models parameters, and, $(b)$ the impact of long-range dependent user/items interactions for prediction. Thereafter, we turn this preliminary study into a novel and successful strategy combining sequential learning per blocks of interactions and removing user with non--homogeneous and non persistent behavior from the training. Even if simple, our approach proves to be surprisingly effective, with respect to state-of-art approaches, especially when dealing with large scale datasets.

The remainder of the paper is organized as follows. In Section~\ref{s:memory}, we present the mathematical framework, used to model persistence in RS data. Thereafter, we present our novel strategy combining the efficiency of sequential learning per block of interactions and the knowledge of the memory behavior of each user in Section~\ref{s:MASAROS}. We then illustrate that memory is intrinsically present in RS user/items interactions in Section~\ref{s:expe} and prove through extensive experiments the effectiveness of our approach.

\section{A Memory Aware sequential learning strategy for RS}\label{s:framework}

\subsection{Framework}\label{s:memory}
Our claim is that all user/items interactions are not equally relevant in the learning process. We prove in the sequel that we can improve the learning process, considering only the subset of users whose interactions with the system possess two main characteristics : {\it homogeneity in time} and {\it persistency}. The user feedback is {\it homogeneous } in time if it is statistically the same, whatever the time period is. Moreover, the behavior of a given user has to be {\it persistent} through time, that is the current choices are dependent on the whole interaction history of the individual.

\noindent We propose to model these two natural characteristics of user feedbacks, using two mathematical notions introduced for sequential data analysis : {\it stationarity} and {\it long memory}. We recall that a time series $X=\{X_t,t\in\mathbb{Z}\}$, here the sequence of user's feedback, is said to be (wide-sense) stationary (see Section 2.4 in~\cite{brillinger2001time}) if its two first orders moments are homogeneous with time: 
\begin{equation}
\forall t,k,l\in\mathbb{Z},\,\mathbb{E}[X_t]=\mu, \mbox{ and }Cov(X_k,X_l)=Cov(X_{k+t},X_{l+t})
\end{equation} 
Under such assumptions the autocovariance of a stationary process only depends on the difference between the terms of the series $h=k-l$. We set $\gamma(h)=Cov(X_0,X_h)$.\\
The concept of long-range dependence arouses in time series analysis to model memory that can be inherently present in sequential data. We shall apply this concept to study the persistent nature of sequential interactions with the RS of each given user. 

A weakly time series is said to be long-range dependent if for some $d\in (0,1/2)$:
\begin{equation}
\gamma(h)\sim h^{2d-1}\mbox{ as }h\to \infty
\end{equation}
The parameter $d$ is called the memory parameter of $X$ and provides a quantitative measure of the persistence of information related to the history of the time series in the long-run. In the perspective of the inference of the memory parameter $d$, we shall use the alternative definition based on the so-called spectral density. The spectral density is the discrete Fourier transform of the autocovariance function: \begin{equation}
    f(\lambda)=\frac{1}{2\pi}\sum\limits_{h=-\infty}^{+\infty}\gamma(h)e^{-ih\lambda}, \qquad \lambda\in(-\pi,\pi].
\end{equation}
and reflects the energy contains at each frequency $\lambda$ if the times series. Under suitable conditions, a stationary time series $X$ admits memory parameter $d\in (0,1/2)$ iff its spectral density satisfies : \begin{equation}
    f(\lambda)\sim \lambda^{-2d}\mbox{ as } \lambda\to 0\;.
\end{equation} 

In order to infer this memory parameter, we use one of the most classical estimators of the memory parameter, the GPH estimator introduced in~\cite{GPH}. It consists of a least square regression of the log-periodogram of $X$. One first defines a biased estimator of the spectral density function, the periodogram $I(\lambda)$ and evaluate it on the Fourier frequencies $\lambda_k=\frac{2\pi k}{N}$ where $N$ is the length of the sample : \begin{equation}
    I(\lambda_k)=\frac{1}{N}\left|\sum\limits_{t=1}^N X_t e^{it\lambda_k}\right|^2
\end{equation}

The estimator of the memory parameter is therefore as follow : \begin{equation}
    \hat{d}(m)=\frac{\sum_{k=1}^m(Y_k-\Bar{Y})\log(I(\lambda_k))}{\sum_{k=1}^m(Y_k-\Bar{Y})^2},
\end{equation}

 where $Y_k=-2\log|1-e^{i\lambda_k}|$, $\Bar{Y}=(\sum_{k=1}^m Y_k)/m$ and $m$ is the number of used frequencies.
\subsection{Learning scheme}\label{s:MASAROS}
We now present our learning scheme. Here, the aim is to take into account the sequence of negative feedback along with positive ones for learning, and select users characteristics as stationarity and long memory. In the following, we first present our Sequential learning with Negative And Positive fEedback approach (called \SO) and then detail the explicit inclusion of memory in the algorithm (that we refer to as \MOSAIC).

\begin{figure}
    \centering
    \includegraphics[width=\textwidth]{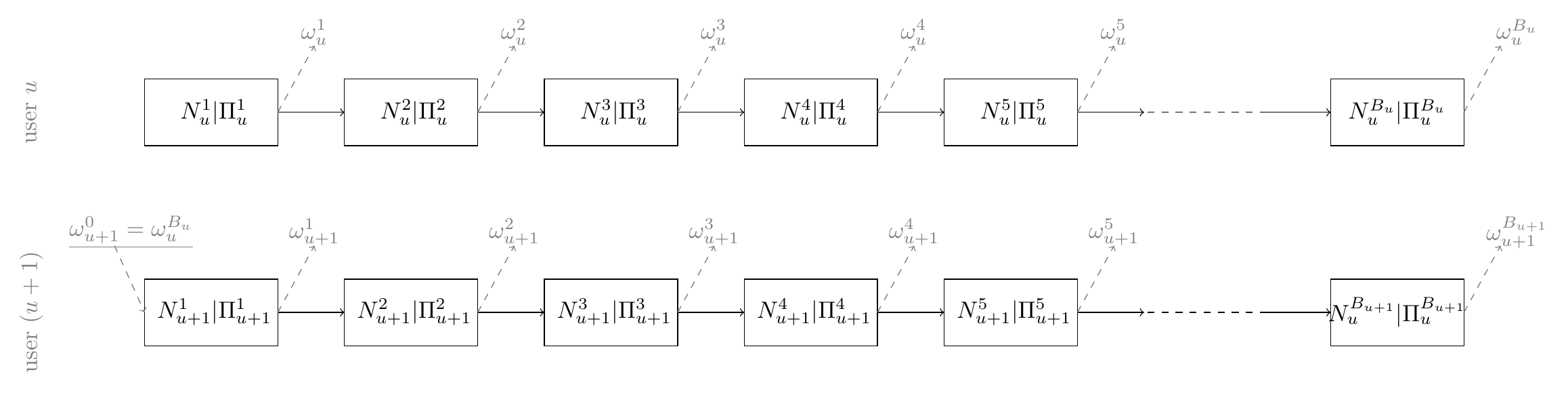}
    \caption{The horizontal axis represents the sequence of interactions over items ordered by time. Each update of weights $\omega_u^{t}; t\in\{b,\ldots,B\}$ occurs whenever the corresponding sets of negative interactions, $\text{N}^t_u$, and positive ones, $\Pi_u^t$, exist.}
    \label{fig:SILICOM}
\end{figure}

User preference over items depend mostly on the context where these items are shown to the user. A user may prefer (or not) two items independently one from another, but within a given set of shown items, he or she may completely have a different preference over these items. By  randomly sampling triplets constituted by a user and corresponding clicked and unclicked items selected over the whole set of shown items to the user, this effect of local preference is not taken into account. Furthermore, triplets corresponding to different users are non uniformly distributed, as interactions vary from one user to another user, and for parameter updates; triplets corresponding to low interactions have a small chance to be chosen. In order to tackle these points; we propose to update the parameters sequentially over the blocks of non-preferred items followed by preferred ones for each user $u$. The constitution of $B+1$ sequences of non-preferred and preferred blocks of items for two users $u$ and $u+1$ is shown in Figure \ref{fig:SILICOM}.

In this case, at each time  a block $\mathcal{B}_u^t=\text{N}_u^{t}\sqcup\Pi_u^{t}$ is formed for user $u$. 

In a classical way~\cite{hu2008collaborative}, each user $u$ and each item $i$ are represented respectively by low dimensional vectors $U_u$ and $V_i$ living in the same latent space of dimension $k$. The goal of the sequential part of our algorithm is to learn a relevant representation of the couples users/items $\omega=(U,V)$ where $U=(U_u)$, $V=(V_i)$. Weights are updated by minimizing the ranking loss corresponding to this block~:
\begin{equation}
\label{eq:CLoss}
    {\hat {L}}_{\mathcal{B}_u^t}(\omega_u^{t}) = \frac{1}{|\Pi_u^{t}||\text{N}_u^{t}|}\sum_{i \in \Pi_u^{t}} \sum_{i'\in \text{N}_u^{t}} \ell_{u, i, i'} ({\omega}_u^{t})\;,
\end{equation}
where $\ell_{u, i, i'}$ is an instantaneous loss, which may be for example the regularized version of the logistic one 
\[
\ell_{u, i, i'}=\log\left(1+e^{-y_{u,i,i'}U_u(V_i-V_{i'})}\right)+\lambda \left(\|U_u\|^2_2+\|V_i\|^2_2+\|V_{i'}\|^2_2\right)
\]
with $y_{u,i,i'}=1$ if the user $u$ prefers item $i$ over item $i'$, $y_{u,i,i'}=-1$ otherwise.

We now describe the inclusion of the Memory-Aware step of our algorithm, allowing to include stationarity and long-memory in the pipeline (called \MOSAIC). In the first step we train \SO{} on the full dataset. Thereafter we remove non stationary embeddings, then test long-range dependence on the remainder and remove non-long range dependent embeddings. Finally we train \SO{} on the filtered dataset. 

The pseudo-code of our Memory-Aware Sequential Learning approach is shown in the following. 

\begin{algorithm}[ht]
   \caption{MemOry-aware Sequential leArning for Implicit feedbaCk (\MOSAIC)}
   \label{alg:CC-conv}
\begin{algorithmic}
\State {\bfseries Input:} A sequence (user and items)  $\{(u,(i_1, \dots,i_{|\mathcal{I}_u|})\}_{u=1}^N$ containing sequences of LRD interactions with the system ; 
\State {\bfseries Preprocessing:} Train the Sequential Learning approach (\SO) on the full dataset.
\State {\bfseries Memory-Aware Step:} Test stationarity for each user. \\
Remove non stationary embedding.\\
Test long-range dependence on the remaining dataset\\
Remove non long-range dependent embeddings
\State {\bfseries Postprocessing:} Train \SO{} on the filtered dataset.
\State {\bfseries Return:} The last updated weights; 
\end{algorithmic}
\end{algorithm}
\section{Experiments and results}\label{s:expe}
In this section, we provide an empirical evaluation of our approach on some popular benchmarks
proposed for evaluating RS. All subsequently discussed components were implemented in Python3 using
the TensorFlow library. We first proceed with a presentation of the general experimental setup, including
a description of the datasets and the baseline models.

\subsection{Datasets}
\noindent{\bf Description of the datasets.} We have considered the three following publicly available datasets, for the task of personalized Top--N recommendation:
\begin{itemize}
    \item ML--1M~\cite{harper2015movielens} consists of user-movie
ratings, on a scale of one to five, collected from a movie recommendation
service. We consider ratings greater
or equal to 4 as positive feedback, and negative feedback otherwise.
\item Kasandr dataset~\cite{sidana2017kasandr} contains 15,844,717 interactions of more than 2 million users in Germany using Kelkoo’s online advertising platform (\url{https://www.kelkoogroup.com/}).
\item Pandor~\cite{sidana2018learning} is another publicly available dataset for online recommendation provided by Purch (\url{http://www.purch.com/}). The dataset records 2,073,379 clicks generated by 177,366 users of one of the Purch’s high-tech website over 9,077 ads they have been
shown during one month.
\end{itemize}

Tables \ref{tab:datasets} presents some detailed statistics about the datasets and the blocks for each collection. 

\begin{table}[ht!]
    \centering
    \begin{adjustbox}{width=\textwidth,center}
    \begin{tabular}{lcccccc}
    \hline
    Data&$|U|$ & $|MEM_{emb}\_U|$ & $|I|$ & Sparsity & Avg. \# of $+$ & Avg. \# of $-$\\
    \hline
    ML-1M&6,040&3,315&3,706&.9553&95.27& 70.46\\
    Kassandr&2,158,859& 2,940 &291,485&.9999&2.42& 51.93\\
    Pandor&177,366&2,153&9,077&.9987&1.32& 10.36\\
    \hline
    \end{tabular}
    \end{adjustbox}
    \caption{Statistics on the \# of users and items; as well as the sparsity and the average number of $+$ (preferred) and $-$ (non-preferred) items on ML-1M, Kassandr and Pandor collections.}
    \label{tab:datasets}
\end{table}

Among these, we report the number of users, $|U|$, and items, $|I|$, the remaining number of users after filtering based on memory in embeddings, $|MEM_{emb}\_U|$, and the average numbers of positive (clicks) and negative feedback (viewed but not clicked). 

\begin{table}[ht!]
    \centering
    \begin{tabular}{lcccc}
    \hline
    Dataset&$|S_{train}|$~~~~&~~~$|S_{test}|$~~&~~$pos_{train}$~~&~~$pos_{test}$\\
    \hline
    {\ML}-1M&797,758&202,451&58.82&52.39\\
    \kasandr&12,509,509&3,335,208&3.36&8.56\\
    Pandor&1,579,716&493,663&11.04&12.33\\
    \hline
    \end{tabular}
    \caption{Number of interactions used for train and test on each dataset, and the percentage of positive feedback among these interactions.}
    \label{tab:detail_setting}
\end{table}

We keep the same set of users in both train and test sets. For training, we use the 80\% oldest interactions of users and the aim is to predict the 20\% most recent user interactions. Table \ref{tab:detail_setting}, presents the number of interactions in train and test as well as the percentage of clicks (positive feedback) in these two sets.

{\bf Identifying stationary and LRD users.} We now describe our filtered subset containing only users with stationary and persistent behavior. Filtering with respect to these two criteria users will allow us to select an interesting subset of the training set, with better generalization properties as shown in next section.

We keep only users whose embeddings have four stationary components, and thereafter keep those whose embeddings components are both LRD. The output subset obtained is much more small for Kassandr and Pandor that the full dataset. 

\begin{table}[ht]
    \centering
    \begin{tabular}{lcccccc}
    \hline
    Data&Total & Stationary & Stationary and LRD  \\
    \hline
    ML-1M&6,040&2,509&2,102\\
    Kassandr&282,500&16,584& 2,029\\
    Pandor&177,366&4,089&1,721\\
    \hline
    \end{tabular}
    \caption{Statistics on the number of total, stationary, stationary and LRD sequences of embeddings.}
    \label{tab:datasets-emb}
\end{table}
The histogram of the memory parameter of sequences of embeddings having four stationary components is given for \ML-1M and \kasandr{} just below. We see clearly that all the components of the embeddings have the same distribution, meaning that all components are equivalent. In addition, we recover that the distribution of user with memory interactions is very different in these two datasets, with a median memory parameter around $0.2$ for \ML-1M and $0.0$ for \kasandr{}. The effect of the Memory-Aware step is then expected to be very different in these two cases.

\begin{figure}[!ht]
    \includegraphics[width=\textwidth]{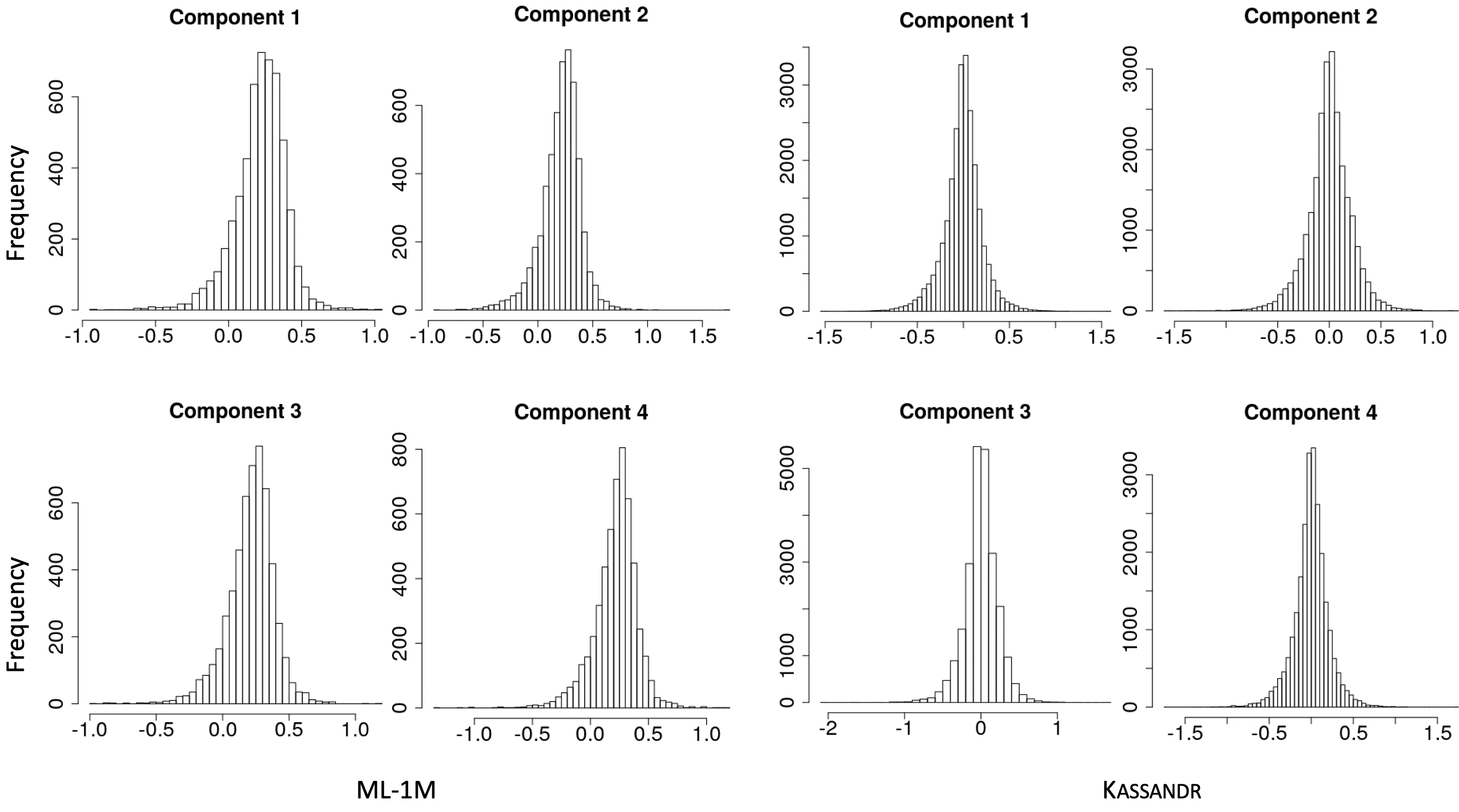}
\caption{Distribution of memory parameter for sequence of embeddings with respect to frequency for \ML-1M (left) and \kasandr{} (right).}
\end{figure}

\subsection{\bf Evaluation}

We consider the following classical metrics for the comparison of the models.
The Mean Average Precision at rank $K$ (MAP@K) over all users  defined as $MAP@K=\frac{1}{N}\sum_{u=1}^{N}AP_K(u)$, where $AP_K(u)$ is the average precision of  preferred items of user $u$ in the top $K$ ranked ones. The Normalized Discounted Cumulative Gain at rank $K$ that computes the ratio of the obtained ranking to the ideal case and allow to consider not only binary relevance as in Mean Average Precision, $NDCG@K = \frac{1}{N}\sum_{u=1}^{N}\frac{DCG@K(u)}{IDCG@K(u)}$, where $DCG@K(u)=\sum_{i=1}^{K}\frac{2^{rel_{i}}-1}{\log_{2}(1+i)}$, $rel_{i}$ is the graded relevance of the item at position $i$; and $IDCG@K(u)$ is $DCG@K(u)$ with an ideal ordering equals to $\sum_{i=1}^{K}\frac{1}{\log_{2}(1+i)}$.

\subsubsection{Models}
To validate our approach described in the previous sections, we compared \SO{} and \MOSAIC{}\footnote{The source code will be made available for research purpose.} with the following approaches. 

{\BPR} \cite{rendle_09} corresponds to a stochastic  gradient-descent  algorithm, based  on  bootstrap  sampling  of  training  triplets, for finding  the model parameters $\omega=(U,V)$ by minimizing the ranking loss over all the set of triplets simultaneously (without considering the sequence of interactions). Although a few extensions of {\BPR} are available \cite{burashnikova2021learning,burashnikovasequential}, {\BPR} remains the golden standard in this area. 
\GRU{} \cite{hidasi2018recurrent} is an extended version of GRU for session-based recommendation. The approach considers the session as the sequence of clicks of the user and learns  model parameters by optimizing a regularized approximation of the relative rank of the relevant item. {\caser} \cite{tang2018caser} is a CNN based model that embeds a sequence of clicked items into a temporal image and latent spaces and find local characteristics of the temporal image using convolution filters.

Hyper-parameters of different models and the dimension of the embedded space for the representation of users and items; as well as the regularisation parameter over the norms of the embeddings for all approaches were found by cross-validation. 

\begin{table}[t]
    \centering
     {
     \begin{tabular}{c|ccc|ccc}
    \hline
      &\multicolumn{3}{c|}{\mapfive}&\multicolumn{3}{c}{\mapten}\\
     \cline{2-7}
     &{\ML-1M}&{\kasandr}&{Pandor}&{\ML-1M}&{\kasandr}&{Pandor}\\
     \hline
     \BPR{}  & 0.852 & 0.316 & 0.431 & 0.819 & 0.337 & 0.448 \\
     \caser  & 0.701 & 0.136 & 0.201 & 0.677 & 0.149 & 0.224\\
     \GRU & 0.761 & 0.636 & 0.318 & 0.736 & 0.637 & 0.332\\
     \SO &{\bf 0.854} & {\bf 0.730} & {\bf 0.457}  & {\bf 0.821} & {\bf 0.730} & {\bf 0.472}\\
     \hline
    \end{tabular}
    }
~\\
~\\
     {\begin{tabular}{c|ccc|ccc}
    \hline
    \cline{2-7}
     &\multicolumn{3}{c|}{\ndcgfive}&\multicolumn{3}{c}{\ndcgten}\\
     \cline{2-7}
     &{\ML-1M}&{\kasandr}& Pandor
     &{\ML-1M}&{\kasandr}&Pandor\\
     \hline
     \BPR{}  & 0.787 & 0.369 & 0.532 & 0.882 & 0.440 &  0.621 \\
     \caser & 0.643 & 0.179 & 0.285 & 0.769 & 0.220 & 0.358\\
     \GRU & 0.709 & 0.670 & 0.397 & 0.822 & 0.693 & 0.489\\
    \SO  & {\bf 0.789} & {\bf 0.754} & {\bf 0.552}  & {\bf 0.885} & {\bf 0.780} & {\bf 0.634}  \\
    
     \hline
    \end{tabular}
    }
    \caption{Comparison of different models in terms of \mapfive{} and \mapten (top), and \ndcgfive{} and \ndcgten (down). }
    \label{tab:measures_state_of_art}
\end{table}

Table~\ref{tab:measures_state_of_art} presents  the comparison of \BPR{}, {\caser} and Sequential Learning approaches over the logistic ranking loss. These results suggest that sequential learning with both positive and negative feedback as it is considered in \SO{} is effective compared to \BPR{} which does not model the sequence of interactions, and {\caser} and \GRU{} which only consider the positive feedback.
\subsubsection{Evaluation of the Memory-Aware Step}

We compare now \SO{} and its enhanced version, adding the Memory-Aware step (\MOSAIC).

Table~\ref{tab:measures_estimation_emb} presents {\mapfive} and {\mapten} (top), and {\ndcgfive} and {\ndcgten} (down)  of  the approaches over the test sets of the different collections. We put in boldface for each dataset the best model. To be able to compare the two approaches, the test set is always the same for \SO{} and the memory-aware one (\MOSAIC). 

On both metrics, {\MOSAIC} outperforms {\SO} on the two large scale datasets, \kasandr{} and Pandor, whereas for \ML-1M the performance are similar. 

What is noticeable, is that the increasing of performance for \kasandr{} and Pandor, is correlated with the removal of nearly 99\% of the users for \kasandr{} and Pandor in the filtering step. It shows that for large scale datasets the filtered subset has better generalisation ability than the full dataset and that memory helps in the learning process. The consistency of the behavior of stationary and LRD users makes the sequence of their interactions much predictable than those of generic users, who may be erratic in their feedback and add noise in the dataset. 

\begin{table}[t]
    \centering
     {
     \begin{tabular}{c|ccc|ccc}
    \hline
      &\multicolumn{3}{c|}{\mapfive}&\multicolumn{3}{c}{\mapten}\\
     \cline{2-7}
     &{\ML-1M}&{\kasandr}&{Pandor}&{\ML-1M}&{\kasandr}&{Pandor}\\
     \hline
     \SO &{\bf 0.854} & 0.730 & 0.457  & {\bf 0.821} & 0.730 & 0.472 \\
      \MOSAIC& 0.839 & {\bf 0.781} & {\bf 0.499} & 0.808 & {\bf 0.780} & {\bf 0.509}\\
     \hline
    \end{tabular}
    }
~\\
~\\
     {\begin{tabular}{c|ccc|ccc}
    \hline
    \cline{2-7}
     &\multicolumn{3}{c|}{\ndcgfive}&\multicolumn{3}{c}{\ndcgten}\\
     \cline{2-7}
     &{\ML-1M}&{\kasandr}& Pandor
     &{\ML-1M}&{\kasandr}&Pandor\\
     \hline
     \SO  & {\bf 0.789} & 0.754 & 0.552  & {\bf 0.885} & 0.780 & 0.634  \\
     \MOSAIC & 0.774 &  {\bf 0.786} & {\bf 0.567} &  0.875 & {\bf 0.809} & {\bf 0.647} \\
     \hline
    \end{tabular}
    }
    \caption{Comparison of \SO{} and \MOSAIC{} in terms of \mapfive{} and \mapten (top), and \ndcgfive{} and \ndcgten (down). }
    \label{tab:measures_estimation_emb}
\end{table}
\section{Conclusion}
We proposed a way to take into account implicit feedback in recommender systems. In addition, we introduce a strategy to filter the dataset with respect to homogeneity and persistency of the behavior in the users when interacting with the system. Surprisingly, when training from this much smaller filtered dataset, the performance of the RS highly improves with respect to several classical metrics showing that we selected users with highly generalisable prediction properties.  
\bibliography{references}

\begin{thebibliography}{10}

\bibitem{brillinger2001time}
D.~R. Brillinger.
\newblock {\em Time series: data analysis and theory}.
\newblock {SIAM}, 2001.

\bibitem{burashnikovasequential}
A.~Burashnikova, Y.~Maximov, and M.-R. Amini.
\newblock Sequential learning over implicit feedback for robust large-scale
  recommender systems.
\newblock In {\em {ECML PKDD}: Joint European Conference on Machine Learning
  and Knowledge Discovery in Databases}, pages 253--269, 2019.

\bibitem{burashnikova2021learning}
A.~Burashnikova, Y.~Maximov, M.~Clausel, C.~Laclau, F.~Iutzeler, and M.-R.
  Amini.
\newblock Learning over no-preferred and preferred sequence of items for robust
  recommendation.
\newblock {\em Journal of Artificial Intelligence Research}, 71:121--142, 2021.

\bibitem{donkers2017sequential}
T.~Donkers, B.~Loepp, and J.~Ziegler.
\newblock Sequential user-based recurrent neural network recommendations.
\newblock In {\em Proceedings of the Eleventh {ACM} Conference on Recommender
  Systems}, pages 152--160, 2017.

\bibitem{Fang:20}
H.~Fang, D.~Zhang, Y.~Shu, and G.~Guo.
\newblock Deep learning for sequential recommendation: Algorithms, influential
  factors, and evaluations.
\newblock {\em {ACM} Transactions on Information Systems}, 39(1), 2020.

\bibitem{GPH}
J.~Geweke and S.~Porter-Hudak.
\newblock The estimation and application of long memory time series models.
\newblock {\em Journal of Time Series Analysis}, 4(4):221--238, 1983.

\bibitem{harper2015movielens}
F.~M. Harper and J.~A. Konstan.
\newblock The movielens datasets: History and context.
\newblock {\em {ACM} transactions on interactive intelligent systems},
  5(4):1--19, 2015.

\bibitem{hidasi2018recurrent}
B.~Hidasi and A.~Karatzoglou.
\newblock Recurrent neural networks with top-k gains for session-based
  recommendations.
\newblock In {\em 27th {ACM} International Conference on Information and
  Knowledge Management ({CIKM})}, pages 843--852, 2018.

\bibitem{hu2008collaborative}
Y.~Hu, Y.~Koren, and C.~Volinsky.
\newblock Collaborative filtering for implicit feedback datasets.
\newblock In {\em International Conference on Data Mining}, pages 263--272.
  {IEEE}, 2008.

\bibitem{rendle_09}
S.~Rendle, C.~Freudenthaler, Z.~Gantner, and L.~Schmidt-Thieme.
\newblock {BPR}: {B}ayesian personalized ranking from implicit feedback.
\newblock In {\em Proceedings of the 25th Conference on Uncertainty in
  Artificial Intelligence ({UAI})}, pages 452--461, 2009.

\bibitem{sidana2018learning}
S.~Sidana, C.~Laclau, and M.-R. Amini.
\newblock Learning to recommend diverse items over implicit feedback on
  {PANDOR}.
\newblock In {\em Proceedings of the 12th {ACM} Conference on Recommender
  Systems}, pages 427--431, 2018.

\bibitem{sidana2017kasandr}
S.~Sidana, C.~Laclau, M.~R. Amini, G.~Vandelle, and A.~Bois-Crettez.
\newblock {KASANDR:} a large-scale dataset with implicit feedback for
  recommendation.
\newblock In {\em Proceedings of the 40th International {ACM} {SIGIR}
  Conference on Research and Development in Information Retrieval}, pages
  1245--1248, 2017.

\bibitem{tang2018caser}
J.~Tang and K.~Wang.
\newblock Personalized top-{N} sequential recommendation via convolutional
  sequence embedding.
\newblock In {\em Proceedings of the 11th {ACM} International Conference on Web
  Search and Data Mining {(WSDM)}}, pages 565--573, 2018.

\bibitem{zhang2019deep}
S.~Zhang, L.~Yao, A.~Sun, and Y.~Tay.
\newblock Deep learning based recommender system: A survey and new
  perspectives.
\newblock {\em {ACM} Computing Surveys ({CSUR})}, 52(1):1--38, 2019.

\end{thebibliography}
\end{document}